\documentclass[prd,preprint,a4paper,superscriptaddress,nofootinbib,11pt]{revtex4}
\usepackage{amsmath,amsfonts,amssymb}
\usepackage{txfonts}
\usepackage{epsfig}
\usepackage{graphics}
\usepackage[T1]{fontenc}
\usepackage[safe]{textcomp}
\usepackage{dsfont}

\usepackage{csquotes}

\begin{document}

\begin{center}

{\bf  \Large Diffusion in kappa deformed space and Spectral Dimension }
 
\bigskip

\bigskip
Anjana V. {\footnote{e-mail: anjanaganga@gmail.com}} \\
School of Physics, University of Hyderabad,
Central University P O, Hyderabad-500046,
India \\[3mm] 
\end{center}

\setcounter{page}{1}
\bigskip

\begin{center}
 {\bf Abstract}
\end{center} 
In this paper, we derive the expression for spectral dimension using a modified diffusion equation in the kappa-deformed space-time. We start with the Beltrami-Laplace operator in the kappa-Minkowski space-time and obtain the deformed diffusion equation. From the solution of this deformed diffusion equation, we calculate the spectral dimension which depends on the deformation parameter `$a$' and also on an integer `$k$', apart from the topological dimension. Using this, we show that, for large diffusion times the spectral dimension approaches the usual topological dimension where as spectral dimension diverges to $+\infty$ for $k\geq 0$ and $-\infty$ for $k < 0$ at high energies . 
\newpage
 
\section{Introduction}
The dimension is a fundamental characteristic property of the space that describes the degrees of freedom experienced by a particle living in that space. Recent works trying to unravel the quantum structure of the space-time have accumulated evidences suggesting that the dimension of space-time is a function of the probe scale \cite{dimred1, dimred2, dimred3, loop, dim uni, DB, diff minko, LMPN, LM}. Further, it has been shown that the dimension of space-time reduces from its topological dimension in some models, while in certain other models, it increases, at high energies. A useful method to obtain the effective dimension of a space-time is by analysing the diffusion process defined in that space-time. The dimension found using such a method is known as spectral dimension \cite{dimred2}. 

The basic notion of the spectral dimension is as follows. A test particle is allowed to diffuse in the concerned space and by studying the rate at which this particle moves from one point to another, one will be able to measure the effective dimension of the space. For an n-dimensional Euclidean space, the motion of the test particle is governed by the diffusion equation
  \begin{equation}
   \frac{\partial }{\partial \sigma}U(x,x';\sigma) = \mathcal{L} U(x,x';\sigma)
  \end{equation}
where $\mathcal{L}$ is the generalised Laplacian and $U(x,x';\sigma)$ is the heat kernel, which gives the probability density of diffusion of the test particle from $x'$ to $x$ during the diffusion time $\sigma$. In order to obtain an estimate of spectral dimension one needs to introduce the concept of average return probability, and it is defined in the following manner
  \begin{equation}
   P(\sigma)= \frac{\int d^n x \sqrt{det g_{\mu \nu}}U(x,x;\sigma)}{\int d^n x \sqrt{det g_{\mu \nu}}},\label{rp}
  \end{equation}
where $g_{\mu \nu}$ is the metric of the underlying space. Spectral dimension $D_s$ is extracted by taking 
the logarithmic derivative of return probability P($\sigma$), i.e.,
  \begin{equation}
  D_s = -2 \frac{\partial \ln P(\sigma)}{\partial \ln \sigma}.
  \end{equation}

Analysis of spectral dimension plays an important role in investigating the nature of space-time at microscopic level. The study on quantum gravity models such as loop quantum gravity, Causal dynamical triangulation, noncommutative space-time shows that the spectral dimension varies as we move from ultraviolet region to infrared region \cite{loop, dim uni, DB, diff minko, LMPN, LM}. Many works along these lines show that, at high energies the spectral dimension converges to two. This feature gives us a hope of setting up a renormalizable model of quantum gravity. 

One of the neat ways to capture the structure of space-time at Planck scale is by using the idea of noncommutativity \cite{nonc}. We are particularly interested in a special case of noncommutative space-time called $\kappa$-sapce-time \cite{kappa1, kappa2}. $\kappa$-space-time is defined as a space-time with a Lie algebraic type relation between space-time coordinates, which can be written as 
\begin{equation}
 [\hat{x}_i , \hat{x}_j] =0 ,~~~~~~~  [\hat{x}_0 , \hat{x}_i] = ia \hat{x}_i,~~~~~~~i,j=1,2,...,n-1 ,\label{ncalgebra}
\end{equation}
where the deformation parameter has the dimension of length and expected to be of the order of Planck length.

As for other quantum gravity models, the diffusion on this space also shows that the effective dimension is different from the usual topological dimension \cite{DB, diff minko, AH1, AH2}. Many other interesting aspects of $\kappa$-deformed space-time have also been attracting attention in recent times \cite{Realization, twist, Deformed, akk, maxwell, PRD86, TJ1, KSGSM1, KSGSM2, HZ1, HZ2}. 

The spectral dimension of (the Wick-rotated) $\kappa$-Minkowski space was analyzed numerically in \cite{DB}. The trace of the heat kernel was calculated using the Casimir of the $\kappa$-Poincare algebra. It was shown that the spectral dimension changes from 3 to 4 as the probe scale increases. In \cite{diff minko}, the spectral dimension was studied using three different forms of $\kappa$-deformed Laplacian. For the Laplacian associated with bi-covariant differential calculus on $\kappa$-space-time, the spectral dimension decreases from 4 at low energies to 3 at high energies. The diffusion process governed by the bi-crossproduct Casimir of $\kappa$-Poincare algebra shows super-diffusion, i.e., the diffused particle experiences more dimension than the usual topological dimension. In the last case, the Laplacian was constructed using the expression for geodesic distance incorporating the notion of relative locality\cite{RL1, RL2, RL3, RL4}, and it was shown that the spectral dimension diverges to infinity as energy increases. In the above works, return probabilities were calculated using energy-momentum relations written in the momentum space. 

In an earlier work \cite{AH1}, spectral dimensions associated with diffusion processes on $\kappa$-deformed Euclidean space were studied, where deformed heat equation was the starting point of the analysis. The deformed Laplacian, which is the Casimir of the undeformed $\kappa$-Poincare algebra is used to construct the diffusion equation. The spectral dimension obtained using this shows a length scale dependence. For various possible modified diffusion equations, similar behaviour for spectral dimension was shown in \cite{AH2}. In \cite{AH2}, the $\kappa$-deformed diffusion equation was constructed using nonrelativistic limit of the $\kappa$-deformed Klein-Gordon operator. 

In this paper, we calculate the spectral dimension associated with diffusion processes on Wick-rotated $\kappa$-Minkowski space-time. Here the modified diffusion equation is constructed using a Beltrami-Laplace operator, different from the one considered in \cite{AH1}. This Beltrami-Laplace operator is written in terms of commutative coordinates and its derivatives, in the $\kappa$-deformed Minkowski space-time. Using Wick's rotation, we obtain the heat equation in the $\kappa$-deformed Euclidean space. Then we solve this deformed diffusion equation perturbatively, and obtain the solution valid upto first order terms in the deformation parameter $a$. This deformed heat kernel is used to calculate the spectral dimension associated with the $\kappa$-deformed Euclidean space and it is seen that the spectral dimension diverges in the limit of probe length going to zero (i.e., at UV region).

This paper is organized as follows. In the next section, we summarise the realization of $\kappa$-deformed space-time and obtain the form of deformed Laplacian in terms of commutative coordinates. In section 3, we construct the $\kappa$-deformed diffusion equation and calculate the spectral dimension. In the last section, we present our concluding remarks.

\section{Undeformed $\kappa$-Poincare algebra}
In this section, we briefly discuss the realizations of the $\kappa$-deformed space-time developed in \cite{Realization}. Studies of physics on $\kappa$-space-time are done either (i) using noncommutative functions defined on noncommutative space-time \cite{Nonfun1, Nonfun2} or (ii) by mapping functions of noncommutative coordinates to functions of commutative coordinates.  In \cite{Realization}, the second approach is developed by constructing a realization of noncommutative space-time coordinates in terms of commutative coordinates $x_\mu$ and their derivatives $\partial_\mu$.

The coordinates of n-dimensional $\kappa$-deformed Minkowski space-time, $\hat{x_\mu}$ satisfy eqn.(\ref{ncalgebra}). The noncommutative coordinates are expressed in terms of commutative coordinates and their derivatives as   
\begin{eqnarray}
  \hat{x_i} =x_i \varphi(A),
  ~~~~~~~~\hat{x_0} = x_0 \psi (A) + iax_k \partial_k \gamma(A),\label{noncomm}
\end{eqnarray}
where $A= -ia\partial_0$. The commutation relations between the noncommutative coordinates, given in eqn.(\ref{ncalgebra}) gives the relation
\begin{equation}
  \frac{\varphi'}{\varphi}\psi = \gamma -1.
\end{equation}
Here $\varphi'=\frac{d\varphi}{dA}$. In the limit $a \rightarrow 0$, we obtain the boundary conditions. 
\begin{eqnarray}
  \varphi(0) = 1, ~~~~~ \psi(0) = 1 ~~~ and ~~ \gamma (0) = \varphi'(0)+1.
\end{eqnarray} 
The requirement that the generators of the symmetry algebra (expressed in terms of the realization given in eqn.(\ref{noncomm})) satisfy the same defining relations as the Poincare algebra, results in the modification of the generators \cite{Realization, twist, Deformed}. This algebra, known as undeformed $\kappa$-Poincare algebra is defined by
\begin{eqnarray}
  [M_{\mu \nu}, D_{\lambda}] = \eta_{\nu \lambda}D_\mu - \eta_{\mu \lambda}D_\nu, ~~~~~ [D_\mu, D_\nu] = 0,
\end{eqnarray}
\begin{equation}
   [M_{\mu \nu}, M_{\lambda \rho}]= \eta_{\nu \lambda}M_{\mu \rho}-\eta_{\mu \lambda}M_{\nu \rho}-\eta_{\nu \rho}M_{\mu  \lambda}+\eta_{\mu\rho}M_{\nu \lambda}.
\end{equation}
Note that the notion of derivative is generalised in the deformed space-time and these deformed derivatives are called Dirac derivatives. The Dirac derivatives are explicitly given by
\begin{eqnarray}
  D_i = \partial_i\frac{e^-A}{\varphi}, ~~~~~ D_0 = \partial_0 \frac{\sinh A}{A}-ia\nabla^2\frac{e^-A}{2\varphi^2}.
\end{eqnarray}
Different realizations of the undeformed $\kappa$-Poincare algebra are obtained by different, allowed choices of $\varphi(A)$ \cite{Realization, Deformed}. Different choices for $\varphi (A) $ are equivalent to different choices for the $*$-products on deformed space-time \cite{star1, star2}. In this paper we chose $\varphi(A) = e^{-A}$. This choice is related to bi-crossproduct basis \cite{varphi1, varphi2} and used in analysing the modification of central potential in the $\kappa$-space-time \cite{akk}.

The Casimir of this undeformed $\kappa$-Poincare algebra can be expressed as 
\begin{equation}
   D_\mu D^\mu = D_i D_i - D_0 D_0 =\square (1+\frac{a^2}{4} \square ),\label{casimir}
\end{equation}
where
\begin{equation}
  \square = \nabla^2_{n-1} \frac{e^{-A}}{\varphi^2}+\partial_{0}^{2}\frac{2(1-coshA)}{A^2}.\label{box}
\end{equation}
For the realization $\varphi(A) = e^{-A} $, we expand eqn.(\ref{casimir}) in terms of deformation parameter $a$ as 
\begin{equation}
   D_\mu D^\mu = \nabla^2_{n-1}-\partial^2_0-ia\nabla^2_{n-1}\partial_0- a^2 \nabla^2_{n-1}\partial^2_0 +\frac{a^2}{3}\partial^4_0 +\frac{a^2}{4}\nabla^4_{n-1}+...
\end{equation}
Now by setting $ x^0 = -ix^n $, we obtain the Beltrami-Laplace operator in the n-dimensional $\kappa$-Euclidean space, i.e.,
\begin{equation}
   D_\mu D_\mu = \nabla^2_{n-1}+\partial^2_n+a\nabla^2_{n-1}\partial_n+a^2 \nabla^2_{n-1}\partial^2_n +\frac{a^2}{3}\partial^4_n +\frac{a^2}{4}\nabla^4_{n-1}+... \label{1}
\end{equation}
Note that the $\square$-operator also reduces to the usual Laplacian in the commutative limit and has been used to analyse the spectral dimension \cite{AH1}. Expansion of eqn.(\ref{box}) with $\varphi(A) = e^{-A}$ in the Euclidean space is
\begin{equation}
   \square = \nabla^2_{n-1}+\partial^2_n+a\nabla^2_{n-1}\partial_n+\frac{a^2}{2} \nabla^2_{n-1}\partial^2_n +\frac{a^2}{12}\partial^4_n+...\label{2}
\end{equation} 
Note that, upto first order correction in $a$, the Casimir and the $\square$-operator are identical ( see eqn.(\ref{1}) and eqn.(\ref{2}) ).

\section{Deformed Diffusion equation and Spectral Dimension}
In this section, we derive the $\kappa$-deformed diffusion equation, using the $\kappa$-deformed Beltrami-Laplace operator, which is expressed in terms of ordinary commutative coordinates and their derivatives. The heat kernel is obtained using a perturbative method and using this we calculate the spectral dimension of (the Wick-rotated) $\kappa$-Minkowski space-time, valid up to first order in $a$. 

Consider a n-dimensional $\kappa$-deformed Euclidean space. The motion of a diffused particle in this space will be governed by the equation 
\begin{equation}
    \frac{\partial}{\partial \sigma}U(x,x';\sigma) = D_{\mu} D_{\mu} U(x,x';\sigma),\label{dieq}
  \end{equation}
where $D_{\mu} D_{\mu}$ is the Beltrami-Laplace operator in the $\kappa$-Euclidean space. We re-express this Laplacian in terms of commutative coordinates and their derivatives, as given in eqn.(\ref{1}).

By restricting our attention to first non-vanishing corrections due to non-commutativity, we rewrite the diffusion equation in the $\kappa$-deformed Euclidean space as 
\begin{eqnarray}
  \frac{\partial U}{\partial \sigma} = \nabla^2_{n-1} U + \partial_n ^2 U + a\nabla^2_{n-1}\partial_n U  = \nabla_n^2 U + a\nabla^2_{n-1}\partial_n U. \label{ncdiff eqn}
\end{eqnarray}
Here $\nabla_n^2 = \nabla^2_{n-1}+ \partial_n ^2$ is the Laplacian in n-dimensional space. It is easy to see that if we are considering the first order correction terms only, the $\square$-operator also leads to the same diffusion equation as in eqn.(\ref{ncdiff eqn}) . By comparing the above deformed diffusion equation with the usual diffusion equation in Euclidean space, we note that the above equation has an extra term, involving product of temporal and spacial derivatives i.e., $\nabla^2_{n-1}\partial_n $.

The heat kernel $U(x,x';\sigma)$ is obtained by solving this equation perturbatively, i.e., we start with the ansatz solution as a perturbative series in $a$ (note that the deformation parameter is expected to be of the order of Planck length), given by
   \begin{equation}
    U=U_0+aU_1+... \label{PS}
  \end{equation}
Using eqn.(\ref{PS}) in eqn.(\ref{ncdiff eqn}) and equating the terms of same order in $a$, we solve equation for $U$ perturbatively. Zeroth order terms in $a$ gives usual heat equation,
  \begin{equation}
    \frac{\partial}{\partial \sigma} U_0(x,x';\sigma)=\nabla^2_{n-1}U_0(x,x';\sigma)+\partial_n^2U_0(x,x';\sigma).\label{u0eqn}
  \end{equation}
The solution of the above equation is
  \begin{equation}
    U_0(x,x';\sigma)=\frac{1}{(4\pi \sigma)^\frac{n}{2}} e^{-\frac{\mid x-x' \mid^2 }{4\sigma}} .\label{U_0}
  \end{equation}
By equating the first order terms in $a$, we obtain
  \begin{equation}
    \frac{\partial}{\partial\sigma}U_1(x,x';\sigma)=\nabla^2_{n-1} U_1(x,x';\sigma)+\partial_n^2 U_1(x,x';\sigma)+\nabla^2_{n-1}\partial_n U_0(x,x';\sigma). \label{u1eqn}
  \end{equation}
Substituting the solution for $U_0$ from eqn.(\ref{U_0}) in eqn.(\ref{u1eqn}), and after a straight forward simplification we end up with the equation for $U_1$ as
  \begin{equation}
    \frac{\partial}{\partial \sigma}U_1(x,x';\sigma) = \nabla^2_{n-1} U_1(x,x';\sigma) +\partial_n^2U_1(x,x';\sigma)+
    \left[\frac{n-1}{4 \sigma^2}(x_n-x'_n) -\frac{(x_n-x'_n)}{8 \sigma^3}\Sigma_{i=1}^{n-1}(x_i-x'_i)^2\right] \frac{1}{(4\pi\sigma)^{\frac{n}{2}}}e^{-\frac{\mid x-x' \mid^2 }{4\sigma}}.\label{eqn}
 \end{equation}
It is known that for a differential equation which  has the general form 
  \begin{equation}
    \frac{\partial}{\partial \sigma}U_1(X,\sigma)=\nabla^2_n U_1(X,\sigma)+f(X,\sigma),\label{dm}
  \end{equation}
the solution satisfying the initial condition
  \begin{equation}
    U_1(X,0)=g(X),
  \end{equation}
is given by \cite{duhamel}
  \begin{equation}
    U_1(X,\sigma)=\int_{R^n} \Phi (X-Y,\sigma)g(Y)dY + 
    \int_{0}^{\sigma}\int_{R^n} \Phi (X-Y,\sigma-s) f(Y,s) dY ds \label{U_1},
  \end{equation}
where 
  \begin{equation}
    \Phi(X,\sigma)=\frac{1}{(4 \pi \sigma)^{\frac{n}{2}}} e^{-\frac{\mid X \mid^2}{4\sigma}}\label{phi}.
  \end{equation}
Using this, we solve the equation for $U_1$ (eqn.(\ref{eqn})). The initial condition satisfied by the solution of eqn.(\ref{eqn}) is 
\begin{equation}
 U_1(X,0)=g(X) = \delta^n(X),\label{IC}
 \end{equation}
where $X=x-x'$. With this initial condition, we calculate the first term on RHS of eqn.(\ref{U_1}) (defined as $U_{11}$). Thus we get
\begin{equation}
 U_{11}(x,x';\sigma) = \int_{R^n} \Phi (X-Y,\sigma)g(Y)dY = 
 \frac{\alpha}{(4 \pi \sigma)^{\frac{n}{2}}} e^{-\frac{\mid x-x' \mid^2}{4 \sigma}}\label{U_11},
\end{equation}
where $\alpha$ has the dimensions of $L^{-1}$. The second term on RHS of eqn.(\ref{U_1}), $U_{12}$ is obtained as
  \begin{eqnarray}
U_{12}(x,x',\sigma) &=& \int_{0}^{\sigma}\int_{R^n} \Phi (X-Y,\sigma-s) f(Y,s) dY ds \nonumber\\
 &=& \frac{1}{(4 \pi \sigma)^\frac{n}{2}}e^{- \frac{\mid x-x' \mid ^2}{4 \sigma}}\{\left(-\frac{(x_n-x'_n)}{8\sigma^3}\Sigma_{i=1}^{n-1}(x_i-x'_i)^2+\frac{n-1}{4\sigma^2}(x_n-x'_n)\right)[\sigma-\epsilon]  \nonumber\\ &-&\frac{1}{4\sigma^2\sqrt{\sigma\pi}} \left[\Sigma_{i=1}^{n-1}(x_i-x'_i)^2+2(x_n-x'_n)\Sigma_{i=1}^{n-1}(x_i-x'_i)\right]\left[\sigma \tan^{-1}\sqrt{\frac{\sigma}{\epsilon}-1}-\epsilon\sqrt{\frac{\sigma}{\epsilon}-1}\right] \nonumber \\ &-&\frac{1}{\sigma^2 \pi}\Sigma_{i=1}^{n-1}(x_i-x'_i)\left[\sigma \ln(\sigma/\epsilon)-\sigma+\epsilon \right] \nonumber \\&+&\frac{n-1}{2\sqrt{\sigma \pi}}\left[5\tan^{-1}\sqrt{\frac{\sigma}{\epsilon}-1}-(4+\frac{\epsilon}{\sigma})\sqrt{\frac{\sigma}{\epsilon}-1}\right]\} \label{U_12}
  \end{eqnarray}
Using eqns.(\ref{U_0},\ref{U_11}) and eqn.(\ref{U_12}) we find the heat kernel valid upto first order in $a$. Using this in eqn.(\ref{rp}), we obtain the return probability as
\begin{equation}
 P(\sigma)= \frac{1}{(4 \pi \sigma)^{\frac{n}{2}}}\left[1+ a \alpha+ a\frac{n-1}{2\sqrt{\sigma \pi}} \left(5\tan^{-1}\sqrt{\frac{\sigma}{\epsilon}-1}-(4+\frac{\epsilon}{\sigma})\sqrt{\frac{\sigma}{\epsilon}-1}\right) \right].
\end{equation}
By taking the logarithmic derivative of $P(\sigma)$, we evaluate the spectral dimension as
\begin{equation}
  D_s = \frac{n+n a \alpha+a \frac{(n-1)}{2\sqrt{\sigma \pi}}\left[5(n+1)\tan^{-1}\sqrt{\frac{\sigma}{\epsilon}-1}-[4n+(n+3)\frac{\epsilon}{\sigma}]\sqrt{\frac{\sigma}{\epsilon}-1}\right]}{1+ a \alpha+ a \frac{(n-1)}{2\sqrt{\sigma \pi}}\left[5\tan^{-1}\sqrt{\frac{\sigma}{\epsilon}-1}-[4+\frac{\epsilon}{\sigma}]\sqrt{\frac{\sigma}{\epsilon}-1}\right]}.
\end{equation} 
Keeping upto first non-vanishing terms in $a$, we obtain the spectral dimension as 
\begin{equation}
   D_s = n+a \frac{(n-1)}{2\sqrt{\sigma \pi}}\left[5\tan^{-1}\sqrt{\frac{\sigma}{\epsilon}-1}-3\frac{\epsilon}{\sigma}\sqrt{\frac{\sigma}{\epsilon}-1}\right].\label{tan}
\end{equation}
 After taking the limit $\epsilon$ to zero, we obtain spectral dimension of the $\kappa$-deformed space-time as
\begin{equation}
  D_s = n+ \frac{5}{4}(n-1)(2k+1)\sqrt{\pi}\frac{a}{\sqrt{\sigma}},~~~~~ k \in \mathbb{Z} .\label{spdm}
\end{equation}
Note that, we have an extra term in the expression for spectral dimension due to non-commutative nature of the space-time. The correction term is proportional to $\frac{a}{\sqrt{\sigma}}$ and is also dependent on the initial dimension $n$. In the commutative limit ($a \rightarrow 0$) the spectral dimension is same as the topological dimension `n'. Here the integer $k$ arises due to the term $\tan^{-1}\sqrt{\frac{\sigma}{\epsilon}-1}$ in eqn.(\ref{tan}). Limit $\epsilon \rightarrow 0$ gives $\tan^{-1} \pm \infty$ which is $(2k+1)\frac{\pi}{2}$. From the expression for spectral dimension (eqn.(\ref{spdm})), we see that the behaviour of spectral dimension depends on $a$ as well as on $k$.  

\begin{figure}[h]
\caption{ spectral dimension as a function of $\sigma$ with $a=1$, $n=4$ and k=0.}\label{fig1}
 \includegraphics[height=2 in, width=2.5 in]{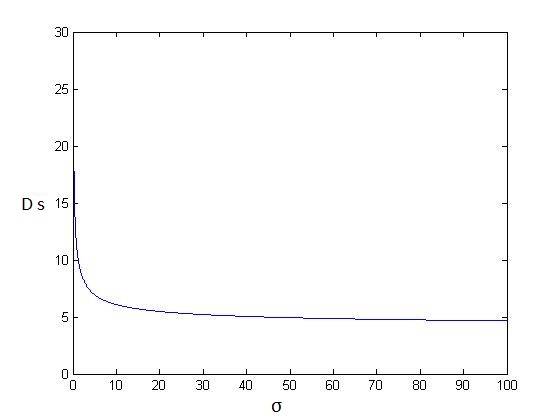}
\end{figure}

The behaviour of the spectral dimension for n=4, k=0 and a=1 is shown in fig.(\ref{fig1}). From this it is easy to see that
\begin{eqnarray*}
  \lim_{\sigma \rightarrow 0} D_s \approx +\infty , ~~~~~~~ \lim_{\sigma \rightarrow \infty} D_s \approx 4.
\end{eqnarray*}
Thus we observe that the dimensions of the space-time increases from the usual topological dimension as $\sigma \rightarrow 0$ and at high energies $D_s \rightarrow \infty$ showing super-diffusion. For all values of $k > 0$, the behaviour of spectral dimension will be same as that of $k = 0$ case.

\begin{figure}[h]
\caption{ spectral dimension as a function of $\sigma$ with $a=1$, $n=4$ and k=-1.}\label{fig2}
 \includegraphics[height=2 in, width=2.5 in]{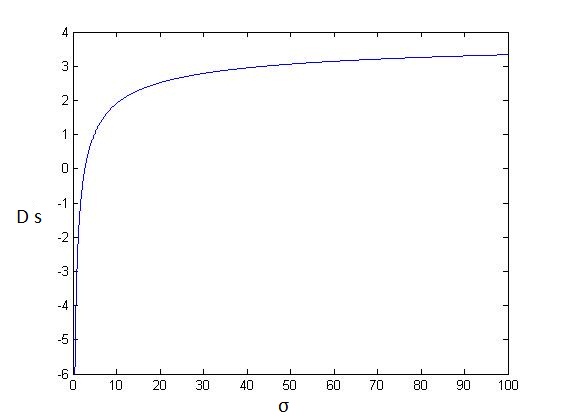}
\end{figure}

For negative values of $k$, the behaviour of spectral dimension is different from what we observed above. The plot for spectral dimension with $k=-1$ is given in fig.(\ref{fig2}) and we see that the spectral dimension flows to $-\infty$ at high energies. In the limit of large diffusion parameter, the spectral dimension is same as the topological dimension.

\section{Conclusion}
In this work, we have constructed the modified diffusion equation for a specific choice of $\varphi$ and calculated the spectral dimension of $\kappa$-deformed space-time. The deformed diffusion equation is derived using the modified Beltrami-Laplace operator which is written in terms of commutative coordinates and their derivatives. The effects of noncommutativity are included through the $a$-dependent terms. Here, the first nonvanishing corrections due to $\kappa$-deformation are in the first order in the deformation parameter. From the analysis, we found that the spectral dimension is changing with the probe scale and it is inversely proportional to $\sqrt{\sigma}$. It also depends on an integer $k$. Here we have seen that, in the small probe scale region, the spectral dimension increases to a value higher than the topological dimension and in the limit $\sigma \rightarrow 0$, $D_s$ goes to infinity for $k\geq 0$. For $k < 0$, the spectral dimension decreases from 4 and reduces to $-\infty$ as $\sigma$ goes to zero. But for the large $\sigma$ values the spectral dimension approaches to the topological dimension in both the cases.  

Behaviour of spectral dimension similar to the $k \geq 0$ case, is reported in \cite{diff minko}. In \cite{diff minko}, the spectral dimension was calculated for three different choices of Laplacian in the momentum space. It was shown that the spectral dimension diverges to infinity as the diffusion time approaches zero for the Laplacian derived using the geodesic distance in the $\kappa$-momentum space. For the Laplacian associated with the bi-cross product Casimir, the spectral dimension varies from 4 to 6 with energy. In \cite{AH1}, it was shown that the spectral dimension diverges to $-\infty$ at UV region as it happens here for $k < 0$. 

In \cite{AH1}, the spectral dimension of $\kappa$-deformed Euclidean space with a different realization ($\varphi(A)=e^{-\frac{A}{2}}$) was studied. Here the deformed diffusion equation was constructed using the Casimir of the undeformed $\kappa$-Poincare algebra and the first nonvanishing correction terms are found to be second order in deformation parameter $a$. It was shown that the spectral dimension decreases from the topological dimension with probe scale. It is known that different choices of realization correspond to different $*$-products \cite{star1, star2} and the realization and $*$-product uniquely chose the vacuum of the theory. Thus, it is natural that the behaviour of spectral dimension is different for the choice $\varphi(A) = e^{\frac{-A}{2}}$ \cite{AH1} and $\varphi(A) = e^A$, as we have seen here.

{\noindent{\bf Acknowledgements}}: The author would like to thank E. Harikumar for suggestions and comments. She also thank University Grants Commission, India, for support through Basic Research Fellowship In Science (BSR) scheme.


\begin{thebibliography}{99}
\bibitem{dimred1} G. 't Hooft, Dimensional Reduction in Quantum Gravity, THU-93/26, arXiv:gr-qc/9310026.
\bibitem{dimred2} J. Ambjorn and Y. Watabiki, Scaling in quantum gravity, Nucl.Phys.B {\bf 445} (1995) 129, arXiv:hep-th/9501049.
 \bibitem{dimred3} S. Carlip, Spontaneous Dimensional Reduction in Short-Distance Quantum Gravity?, AIP Conf. Proc. {\bf 1196} (2009) 72 arXiv:0909.3329.
 \bibitem{loop}L. Modesto, Fractal Structure of Loop Quantum Gravity, Class. Quant. Grav. {\bf 26} (2009) 242002,
arXiv:0812.2214.
\bibitem{dim uni} J. Ambjorn, J. Jurkeiwicz and R. Loll, Spcetral Dimension of the Universe, Phys. Rev. Lett. {\bf 95} (2005) 171301, arXiv:hep-th/0505113. 
\bibitem{DB} D. Benedetti, Fractal properties of quantum space-time, Phys. Rev. Lett. {\bf 102} (2009) 111303,
 arXiv:0811.1396.
 \bibitem{diff minko} M. Arzano and T. Trzesniewski, Diffusion on $\kappa$-Minkowski space, Phys. Rev. D {\bf 89} (2014) 124024, arXiv:1404.4762.
 \bibitem{LMPN} L. Modesto and P. Nicolini, Spectral dimension of a quantum universe, Phys. Rev. D {\bf 81} (2
 010) 104040, arXiv:0912.0220.
\bibitem{LM} L. Modesto, Super-renormalizable quantum gravity, Phys. Rev. D {\bf 86} (2011) 044005, arXiv:hep-th/1107.2403.
 \bibitem{nonc} S. Doplicher, K. Fredenhagen and J. E. Roberts, The quantum structure of spacetime at the Planck scale and quantum fields, Commun. Math. Phys. {\bf 172} (1995) 187, arXiv:hep-th/0303037.
\bibitem{kappa1} J. Lukierski, A. Nowicki and H. Ruegg, New quantum Poincare algebra and $\kappa$-deformed field theory
 Phys. Lett. B {\bf293} (1992) 344.
 \bibitem{kappa2} J. Lukierski and H. Ruegg, Quantum $\kappa$-Poincare in any dimensions, Phys. Lett. B {\bf 329} (1994) 189, arXiv:hep-th/9310117.
\bibitem{AH1} Anjana V. and E. Harikumar, Spectral dimension of $\kappa$-deformed space-time, Phys. Rev. D {\bf 91} (2015) 065026.
\bibitem{AH2} Anjana V. and E. Harikumar, Dimensional flow in the kappa-deformed space-time, Phys. Rev. D {\bf 92} (2015) 045014.
\bibitem{Realization} S. Meljanac and M. Stojic, New realizations of Lie algebra kappa-deformed Euclidean space, Eur. Phys. J. C {\bf 47} (2006) 531, arXiv:0605133.
\bibitem{twist} T. R. Govindarajan, K. S. Gupta, E. Harikumar, S. Meljanac and D. Meljanac, Twisted statistics in $\kappa$-Minkowski spacetime, Phys. Rev. D {\bf 77} (2008) 105010.
\bibitem{Deformed} T. R. Govindarajan, K. S. Gupta, E. Harikumar, S. Meljanac and D. Meljanac, Deformed oscillator algebras and QFT in $\kappa$-Minkowski space-time, Phys. Rev. D {\bf 80} (2009) 025014, arXiv:0903.2355.
\bibitem{akk} E. Harikumar and A. K. Kapoor, Newton's Equation on the kappa space-time and the Kepler problem, Mod. Phys. Lett. A {\bf25} (2010) 2991.
\bibitem{maxwell} E. Harikumar, Maxwell's equations on the $\kappa$-Minkowski spacetime and Electric-Magnetic duality, Euro. Phys. Lett. {\bf 90} (2010) 21001, arXiv:1002.3202.
\bibitem{PRD86} E. Harikumar, A. K. Kapoor and R. Verma, Uniformly accelerating observer in $\kappa$-deformed space-time, Phys. Rev. D {\bf 86} (2012) 045022.
\bibitem{TJ1}T. Juric, S. Meljanac and D. Pikutic, Realizations of $\kappa$-Minkowski space, Drinfeld twists and related symmetry algebras, arXiv:1506.04955.
\bibitem{KSGSM1} K. S. Gupta, S. Meljanac and A. Samsarov, Quantum statistics and noncommutative black holes,  Phys. Rev. D {\bf 85} (2012) 045029, arXiv:1108.0341. 
\bibitem{KSGSM2} S. Meljanac, A. Pachol, A. Samsarov and K. S. Gupta, Different realizations of kappa-momentum space and relative-locality effect, Phys. Rev. D {\bf 87} (2013) 125009. 
\bibitem{HZ1} P. Guha, E. Harikumar and Zuhair N. S., MICZ-Kepler systems in noncommutative space and duality of force laws, Int. J. Mod. Phys. A {\bf 29} (2014) 1450187, arXiv:1404.6321.
\bibitem{HZ2} P. Guha, E. Harikumar and Zuhair N. S., Fradkin-Bacry-Ruegg-Souriau vector in kappa-deformed space-time,  	arXiv:1504.01897.
\bibitem{RL1} G. Amelino-Camelia, L. Freidel, J. Kowalski-Glikman and L. Smolin, The principle of relative locality, Phys. Rev. D {\bf 84} (2011) 084010, arXiv:1101.0931.
\bibitem{RL2}G. Amelino-Camelia, L. Freidel, J. Kowalski-Glikman and L. Smolin, Relative locality: A deepening of the relativity principle, Gen. Relativ. Gravit. {\bf 43} (2011) 2547, arXiv:1106.0313.
\bibitem{RL3} L. Freidel, R. G. Leigh and D. Minic, Born Reciprocity in String Theory and the Nature of Spacetime, Phys. Lett. B {\bf 730} (2014) 302, arXiv:1307.7080.
\bibitem{RL4} L. Freidel, R. G. Leigh and D. Minic, Quantum Gravity, Dynamical Phase Space and String Theory, Int. J. Mod. Phys. D {\bf 23} (2014) 1442006, arXiv:1405.3949.
 \bibitem{Nonfun1} M. Dimitrijevic, L. Jonke, L. Moller, E. Tsouchnika, J. Wess and M. Wohlgenannt, Deformed Field Theory on kappa-spacetime, Eur. Phys. J. C {\bf 31} (2003) 129, arXiv:hep-th/0307149.
 \bibitem{Nonfun2} L. Freidel, J. Kowalski-Glikman and S. Nowak, From noncommutative kappa-Minkowski to Minkowski space-time,  	Phys. Lett. B {\bf 648} (2007) 70, arXiv:hep-th/0612170.
 \bibitem{star1} T. Juric, S. Meljanac and R. Strajn, Twists, realizations and Hopf algebroid structure of kappa-deformed phase space, Int. J. Mod. Phys. A {\bf 29} (2014) 1450022, arXiv:1305.3088.
 \bibitem{star2} K. S. Gupta, E. Harikumar, T. Juric, S. Meljanac and A. Samsarov, Noncommutative scalar quasinormal modes and
quantization of entropy of a BTZ black hole, JHEP {\bf 09} (2015) 025.
\bibitem{varphi1} S. Majid and H. Ruegg, Bicrossproduct structure of $\kappa$-Poincare group and non-commutative geometry, Phys. Lett. B {\bf 334} (1994) 348, arXiv:hep-th/9405107.
\bibitem{varphi2} J. Lukierski, H. Ruegg and W. J. Zakrzewski, Classical and Quantum Mechanics of Free \k Relativistic Systems, Ann. Phys. {\bf 243} (1995) 90, arXiv:hep-th/9312153.
\bibitem{duhamel} F. John, Partial Differential Equations, New York, Springer-Verlag, 1982, 4th ed.
  
\end{thebibliography}
\end{document}